\documentstyle[aps,prd]{revtex}
\tighten
\renewcommand{\epsilon}{\varepsilon}

\begin{document}

\title{Reheating and causal thermodynamics}

\author{Winfried Zimdahl\footnote{Electronic address:  
winfried.zimdahl@uni-konstanz.de}}
\address{Fakult\"at f\"ur Physik, Universit\"at Konstanz, PF 5560 M678
D-78434 Konstanz, Germany
}
\author{Diego Pav\'{o}n\footnote{Electronic address:  
diego@ulises.uab.es}}
\address{Departament de F\'{\i}sica,
Universitat Aut\`{o}noma de Barcelona,
E-08193 Bellaterra (Barcelona), Spain}
\author{Roy Maartens\footnote{Electronic address:  
maartensr@cv.port.ac.uk}}
\address{School of Mathematical Studies, Portsmouth University,
Portsmouth PO1 2EG, England\footnote{permanent address}
\\
and Department of Mathematics, University of Natal, Durban 4041, South
Africa}

\date{\today}

\maketitle


\begin{abstract}

The reheating process in inflationary universe models is considered
as an out-of-equilibrium mixture of two interacting and reacting
fluids, and studied within the framework of causal, irreversible
thermodynamics. The evolution of the temperature
and the decay rate as
determined by causal thermodynamics are estimated
at different stages of the process. A simple model is also used
to find the perturbations of the expansion rate, including the
possibility of damped oscillations.

\end{abstract}

\section{Introduction}

The reheating period is a key ingredient in many inflationary models 
of the very early Universe. During reheating,
most of the matter and radiation
of the Universe is created via the decay of the inflaton field,
while the temperature grows by many orders of magnitude.
Traditionally, the idealized picture of
a very quick decay in which the products reach
equilibrium immediately is assumed \cite{KoTu}.
However,
recent quantum field theoretical investigations
\cite{KoLiSta1} - \cite{Fuji}
indicate that the reheating period is characterized by complicated  
nonequilibrium processes,
the main characteristics of which are initial, violent particle
production via parametric resonance (`preheating'),
with a highly nonequilibrium
distribution of the produced particles, subsequently relaxing to
an equilibrium state.

In this paper we aim at obtaining a phenomenological understanding  
of the reheating process within a model of two interacting and
reacting fluids. We regard this approach as
complementary to the above-mentioned
quantum field theoretical studies.
The model involves a fluid with the equations of state for matter
(modelling  scalar field oscillations about the true ground state
in the reheating) that decays into a  relativistic fluid.
The implications of an intermediate decay into other massive  
(bosonic) particles that does not explicitly occur in this model,  
are assumed to be describable with the help of a large, effective  
decay rate of the
initial into the final component of the entire process, leaving the 
detailed microphysical study of this epoch to quantum field  
theoretical investigations which are beyond the scope of the present  
paper.

While these simplifications may appear drastic, they open the  
possibility
of  studying the backreaction of the decay process as a whole on  
the entire cosmological dynamics, including the behaviour of the  
scale factor.
We believe this to be a main advantage of our approach since most  
of the quantum field theoretical calculations do not even take into  
account the expansion of the universe.

A two-fluid description of the expanding Universe is necessarily
dissipative, even if the components are assumed to be intrinsically
perfect fluids.
This is also true for nonreacting fluids (conserved particle
numbers), where the different cooling rates of the subsystems
produce an effective, entropy generating bulk viscous pressure of the 
system
as a whole \cite{UI}, \cite{ZMN}. In fact
conservation of the particle number is only a very
special case, particularly at high energies.
The processes we are interested in are characterized by particle
decay and production.

As has been shown recently, any deviation from detailed balance in  
the decay and inverse decay reactions in the expanding Universe
gives rise to additional bulk pressures \cite{ZPrep}.
In the present case of strong perturbations of the detailed balance  
(the reactions are predominantly one-directional with the inverse
processes largely suppressed), one expects considerable
bulk viscous pressures that characterize the deviations from
equilibrium of
the cosmological fluid as a whole. To describe this
nonequilibrium process, we shall resort to the well-known
Israel-Stewart theory of transport processes \cite{IS} which,
because of its causality and stability properties \cite{HL},
has been repeatedly applied in the cosmological context
-- see \cite{RP}, \cite{m}, \cite{ZPRD} and references therein.

In this theory, generalized
`fluxes', like a bulk pressure, become dynamical degrees
of freedom on
their own and have associated relaxation times.
Our point of view here is to regard the cosmic substratum during
the reheating period as a causal, dissipative fluid, relaxing to
equilibrium. (By ``causal" we mean that dissipative signals
propagate only at subluminal speeds).

Bearing in mind that the Israel-Stewart theory was derived for
small deviations from equilibrium (for attempts to apply it to  
far-from-equilibrium situations see \cite{m}, \cite{ZPRD}), we
point out that in the present context it is the high `preheating'
particle production rate that provides a `creation' contribution
to the `effective' bulk pressure which may be much larger than the  
conventional, thermodynamic, dissipative bulk pressure.
While the latter is considered to be small in the applications of
this paper, the `effective' bulk pressure, that determines the
reheating temperature, is not.

Section II establishes the basic relationships
concerning both particle and energy--momentum non--conservation
in our two-fluid system, and presents an expression for the
evolution of the temperature of the overall fluid.
Section III introduces the causal transport relation for the
thermodynamic, dissipative bulk stress, and deduces the corresponding 
equation for the  evolution of the Hubble parameter.
Section IVA solves
the latter equation at three different stages of
the reheating by assuming a very short relaxation time,
and determines the temperature evolution in each of them.
The qualitative description of the dynamics is determined essentially
by a single thermodynamical quantity, the dissipative contribution
to the speed of sound. If this is larger than the adiabatic
contribution to the sound speed at the beginning of reheating,
we show that the temperature rises rapidly to a maximum
(reheating) temperature, which we estimate. This condition for
rising temperature is equivalent
to a growth in the total particle number density, which is
reasonable in the initial stage of reheating,
when coherent oscillations of the scalar field lead to huge
production of particles.
Thereafter, the temperature falls, but less
rapidly than in the non--dissipative case.
In Section IVB, we construct a simple
model to calculate the perturbations of the expansion rate due to
causal viscous and reactive effects. The model includes the
possibility of damped oscillations in the beginning of reheating.
Finally, section V summarizes our conclusions.

Units have been chosen so that $c = k_{B} = \hbar = 1$.

\section{The two-fluid model}

Let us assume the energy-momentum tensor
$T ^{ik}$ of the cosmic medium splits into  two perfect fluid parts:
\begin{equation}
T ^{ik} = T ^{ik}_{1} + T ^{ik}_{2} \ ,
\label{1}
\end{equation}
with ($A = 1, 2$)
\begin{equation}
T^{ik}_{A} = \rho_{A} u^{i}u^{k}
+ p_{A} h^{ik}\ . \label{2}
\end{equation}
$p_{A}$ is the equilibrium pressure of species $A$.
For simplicity we assume that both components share the same
$4$-velocity $u^{i}$,
with projection tensor
$h^{ik} =g^{ik} + u^{i}u^{k}$.
The particle flow vector $N_{A}^{i}$ of species $A$ is
\begin{equation}
N_{A}^{i} = n_{A}u^{i}\ ,\label{3}
\end{equation}
where $n _{A}$ is the particle number density.
We are interested in situations where neither the particle numbers  
nor the energy-momentum of the components are separately conserved,  
i.e. particle interconversion and exchange of energy and momentum
between the components are admitted.

The balance laws for the particle numbers are
\begin{equation}
N _{A ;i}^{i} = \dot{n}_{A} + \Theta n_{A} = n _{A} \Gamma _{A} \ ,
\label{4}
\end{equation}
where $\Theta \equiv u^{i}_{;i}$ is the fluid expansion and $\Gamma  
_{A}$ is the rate of change of the number of particles of species
$A$.
There is particle production for $\Gamma _{A} > 0$ and particle
decay for
$\Gamma _{A} < 0$. For $\Gamma _{A} = 0$ we have
separate particle number conservation.

Interactions between the fluid components amount to the mutual
exchange of energy and momentum.
Consequently, there will be no local energy-momentum conservation
for the subsystems separately.
Only the energy-momentum tensor of the system as a whole is conserved. 
Denoting the loss- and source-terms in the separate balances by $t  
^{i}_{A}$, we may write
\begin{equation}
T ^{ik}_{A ;k} = - t _{A}^{i} \ ,
\label{5}
\end{equation}
implying
\begin{equation}
\dot{\rho}_{A} + \Theta\left(\rho_{A} + p_{A}\right) = u _{i}t_{A}^{i}
\ ,
\label{6}
\end{equation}
and
\begin{equation}
\left(\rho _{A} + p _{A}\right) \dot{u}^{i} + p _{A ,k}h ^{ik}
= - h ^{i}_{k}t ^{k}_{A} \ .
\label{7}
\end{equation}
All the considerations to follow will be independent of the
specific structure of the $t _{A}^{i}$. In general, there are no
limitations on the strength or the structure of the interaction.

Each component is governed by a separate Gibbs equation:
\begin{equation}
T_{A} \mbox{d} s _{A} = \mbox{d} \left(\frac{\rho _{A}}{n _{A}}\right) 
+ p _{A}\mbox{d}\left(\frac{1}{n _{A}}\right) \ .
\label{8}
\end{equation}
Using (\ref{4}) and (\ref{6}) one finds for the time evolution of
the entropy per particle
\begin{equation}
n _{A} T _{A} \dot{s}_{A} = u _{i} t _{A}^{i}
- \left(\rho _{A} + p _{A}\right) \Gamma _{A} \ .
\label{9}
\end{equation}
With nonvanishing source terms in the balances for $n _{A}$ and
$\rho _{A}$, the rate of change of
entropy per particle is nonzero in general.
Below we shall deal with the special case that the terms on the
right hand side  of (\ref{9}) just cancel.

According to (\ref{5}) the condition of energy-momentum
conservation for the system as a whole,
\begin{equation}
\left(T _{1}^{ik} + T _{2}^{ik}\right)_{;k} = 0 \ ,
\label{10}
\end{equation}
implies
\begin{equation}
t _{1}^{i} = - t _{2}^{i} \ .
\label{11}
\end{equation}
There is no corresponding condition, however, for the particle
number balance as a whole.
Defining the integral particle number density $n$ as
\begin{equation}
n = n _{1} + n _{2} \ ,
\label{12}
\end{equation}
we have
\begin{equation}
\dot{n} + \Theta n = n \Gamma \ ,
\label{13}
\end{equation}
with
\begin{equation}
n \Gamma = n _{1}\Gamma _{1} + n _{2}\Gamma _{2}
\ .
\label{14}
\end{equation}
$\Gamma$ is the rate by which the total particle number $n$ changes. 
We do {\it not} require $\Gamma$ to be zero, since total particle
number conservation is only a very special case, especially at high  
energies.

From now on we assume that the source terms on the right hand side of  
(\ref{9}) cancel among themselves, i.e., that the entropy per
particle of each of the components is preserved.
The particles decay or come into being with a {\it fixed}
entropy $s _{A}$.
This adiabaticity condition
amounts to the assumption that the particles at any stage are
amenable to a perfect fluid description.
With $\dot{s}_{A} = 0$ in (\ref{9}) one has
\begin{equation}
u _{i} t ^{i}_{A} = \left(\rho _{A} + p _{A}\right) \Gamma _{A} \ .
\label{15}
\end{equation}
This relationship establishes a link between the source
terms in (\ref{4}) and (\ref{6}) which originally are independent
quantities.
The simplifying assumption $\dot{s}_{A} = 0$ takes into account the  
circumstance that the production process itself is the main source
of entropy production, while dissipative processes within each of
the separate components are less important. The cosmic fluid as a
whole will be considered, however, as dissipative (see below).

Combining (\ref{11}) and (\ref{15}) one has
\begin{equation}
u _{i} t ^{i}_{1} = \left(\rho _{1} + p _{1}\right) \Gamma _{1}
= - u _{i} t ^{i}_{2} = - \left(\rho _{2} + p _{2}\right) \Gamma _{2}
\ ,
\label{16}
\end{equation}
which provides us with a relation between the rates $\Gamma _{1}$ and 
$\Gamma _{2}$:
\begin{equation}
\Gamma _{2}=-
\left[\frac{\rho _{1} + p _{1}}{\rho _{2} + p _{2}} \right]
\Gamma _{1}
\ .\label{17}
\end{equation}
Use of this relation in (\ref{14}) yields
\begin{equation}
n \Gamma = n _{1}\Gamma _{1} h _{1}
\left[\frac{1}{h _{1}}
- \frac{1}{h _{2}}\right]
\ ,
\label{18}
\end{equation}
where
$h _{A} \equiv \left(\rho _{A} + p _{A}\right)/ n
_{A}$ are the enthalpies per particle.
Total particle number conservation, i.e. $\Gamma = 0$, is only
possible if  $h _{1} = h _{2}$.

As is shown elsewhere \cite{ZMN}, \cite{ZPrep},
a system of two fluids,
each of them perfect on its own, is dissipative in general and may  
be characterized by an energy-momentum tensor
\begin{equation}
T ^{ik} = \rho u ^{i}u ^{k} + \left(p + \pi\right) h ^{ik} \ .
\label{19}
\end{equation}
The equilibrium pressure $p$ of the total system and the
energy density $\rho $ are assumed to obey equations of state
\begin{equation}
p = p\left(n, n _{1}, T\right)
\label{20}
\end{equation}
and
\begin{equation}
\rho = \rho \left(n, n _{1}, T\right)\ ,
\label{21}
\end{equation}
where $T$ is the equilibrium temperature of the system as a whole,
{\it defined} by
(cf. \cite{UI}, \cite{ZMN})
\begin{equation}
\rho_{1}\left(n_{1},T_{1}\right) + \rho_{2}\left(n_{2},T_{2}\right) 
= \rho \left(n, n _{1}, T\right)
\ . \label{22}
\end{equation}
It is worth mentioning that
there does not exist a corresponding relation for  the pressures.
The partial pressures of the components do not add up to the
equilibrium pressure in general.
The difference between the sum
$p _{1}\left(n _{1}, T _{1}\right) + p _{2}\left(n _{2}, T
_{2}\right)$ and the equilibrium pressure $p \left(n, T\right)$
contributes to the thermodynamical, dissipative bulk pressure $\pi $.
A further source of $\pi $ is deviations from detailed
balance, i.e. contributions due to $\Gamma _{A} \neq 0$
\cite{ZPrep}.

The behavior of the equilibrium temperature $T$ of the system as a  
whole is governed by \cite{ZPrep}
\begin{equation}
\frac{\dot{T}}{T} = - \left(\Theta - \Gamma\right)
\frac{\partial_T{p}}{\partial_T{\rho }}
- \frac{\Theta \pi }{T \partial_{T} \rho}
+ \left(\frac{\dot{T}}{T}\right)_{\!*}
\ ,
\label{23}
\end{equation}
where the abbreviations $\partial_T f=\partial f/\partial T$ and
\begin{eqnarray}
\left(\frac{\dot{T}}{T}\right)_{\!*} & \equiv &
- \frac{p \, \Gamma - p _{1}\Gamma _{1} - p _{2}\Gamma _{2}}
{T \, \partial _{T} \rho} \nonumber\\
&=& \frac{\rho _{1} + p _{1}}{T \partial_T \rho}
\left(\frac{n _{2}}{n} p _{1} - \frac{n _{1}}{n} p _{2}\right)
\left[\frac{1}{n _{1}h _{1}}+\frac{1}{n _{2}h _{2}}\right] \Gamma _{1}
\label{24}
\end{eqnarray}
were used.
The term (\ref{24}) takes into account deviations from
classical gas behavior.
It vanishes for $p _{A} = n _{A}T$, i.e. for a mixture of
classical gases.
For a mixture of a classical and a quantum gas, or for a mixture of  
fermions and bosons, it will be nonzero in general.
We shall restrict ourselves to the case of two classical fluids,
i.e. to the case
$(\dot{T}/T)_* = 0$. This simplifying assumption
is in line with the general restriction of our approach concerning
the detailed microphysics, pointed out in the
introduction.

The temperature evolution equation (\ref{23}) suggests that we  
define an `effective' bulk pressure
\begin{equation}
\pi _{_{eff}} \equiv  \pi - \frac{\Gamma }{\Theta }T \partial _{_{T}}p 
=  \pi - \frac{\Gamma }{\Theta } p \ ,
\label{24a}
\end{equation}
so that (\ref{23}) in the classical case may be written as
\begin{equation}
\frac{\dot{T}}{T} = - \Theta \frac{\partial _{_{T}}{p}}
{\partial _{_{T}}{\rho }}
- \frac{\Theta \pi _{_{eff}}}{T \partial _{_{T}}\rho }\ .
\label{24b}
\end{equation}
Our main concern in the following sections will be to study the
influence of the different parts of the entropy producing  
`effective' bulk pressure
$\pi _{_{eff}}$ on the cosmological evolution.

The $\pi _{_{eff}}$-term in (\ref{24b}) describes the deviation  
from the adiabatic temperature behaviour.
It is obvious that any $\pi _{_{eff}} < 0$ on the right-hand side of 
(\ref{24b}) yields a positive, i.e., `reheating' contribution to
$\dot{T}/T$, counteracting the first term on the right-hand side of 
(\ref{24b}) that simply describes the adiabatic cooling due to the
expansion.

As we shall discuss below the nonequilibrium term on the right-hand 
side of
(\ref{24b}) may overcompensate the adiabatic term during the  
initial `preheating' stage.

The effective bulk pressure (\ref{24a}) consists of the  
conventional, thermodynamic, dissipative bulk pressure $\pi $ and a  
creation part
$- p \Gamma / \Theta $.
For $\Gamma > \Theta $, a condition that is expected to be fulfilled 
during the initial stage of reheating, the creation part gives the
dominant contribution to the effective bulk pressure.
Even for a small or vanishing thermodynamic viscous pressure $\pi $ 
the entropy producing nonequilibrium parts on the right-hand side
of the temperature law (\ref{24b}) may overcompensate the adiabatic  
part,
provided  only that the production rate $\Gamma $ is sufficiently high. 
In other words, even the production of particles with an  
equilibrium distribution, equivalent to the possibility of a perfect  
fluid
description, gives rise to `reheating' and entropy production.
We shall consider this to be the dominant part of the entropy
production during preheating.
It is the particle production process itself which is connected
with entropy production, simply through the enlargement of the
phase space.
There are additional entropy producing contributions  due to the
fact that in reality the particles will deviate from equilibrium.
These contributions will be subject to a causal transport equation
in the following section.

\section{The causal evolution equation}

In the Israel-Stewart second order theory
of irreversible processes,
the viscous pressure $\pi$ is a dynamical degree of freedom on its
own \cite{IS}.
Instead of applying the much more involved full Israel-Stewart theory 
we shall restrict ourselves to the so-called {\it truncated}
version \cite{m}, \cite{ZPRD}, \cite{Roy}
of this theory, since the latter already captures the essence of  
noninstantaneous relaxations.
While the full and the truncated theories may disagree significantly 
if applied to far-from-equilibrium situations \cite{m},  
\cite{ZPRD}, \cite{Roy}, they are expected to provide similar  
results near
equilibrium \cite{ZPRD}.
Now, reheating is anything but close to equilibrium. But as already 
mentioned, the main contribution to the entropy production during
the decay process
may be traced back to the second term in the expression (\ref{24a}) 
for the effective bulk pressure.
The thermodynamic, dissipative bulk pressure $\pi $ may be regarded 
as a small perturbation under these circumstances and is supposed
to obey the truncated causal transport equation
\begin{equation}
\pi + \tau\dot{\pi}  =  - \zeta \Theta \ .
\label{25}
\end{equation}
The quantity $\tau$ denotes the relaxation time associated with the 
thermodynamic, disipative  bulk pressure, i.e. the time the system
would take to come
to equilibrium (perfect fluid behaviour)
if the generalized `force' -- in this case
the expansion $\Theta$ -- were suddenly turned off. It may be related 
to the bulk viscosity $\zeta$ by \cite{Roy}
(see also \cite{ZPRD})
\begin{equation}
\frac{\zeta}{\tau} =  c_b^2  (\rho + p)  \ , \quad
c_b^2  \leq 1 -c_s^2 \ ,
\label{26}
\end{equation}
where $c_s$ is the adiabatic sound speed and
$c_b$ is the bulk viscous countribution to the dissipative speed
of sound $v$, given by $v^2=c_s^2+c_b^2$.
Thus
\begin{equation}
\pi + \tau\dot{\pi}  =  - \Theta c_b^2  \rho \gamma \tau
\ .
\label{27}
\end{equation}
Here and throughout we will use the abbreviation
$$\gamma \equiv 1 + {p\over\rho} \ , $$
where $\gamma$ is {\it not} assumed constant.

The Einstein field equations for a spatially-flat
Robertson-Walker spacetime are
\begin{equation}
3H^{2}  = \kappa\rho
\ ,  \label{28}
\end{equation}
where $\kappa$ is Einstein's gravitational constant,
and
\begin{equation}
\dot{H}  = - {\textstyle{1\over2}}\kappa\left(\rho + p + \pi\right)
\ , \label{29}
\end{equation}
where $H ={1\over3} \Theta= \dot{R}/R$
and $R$ is the cosmic scale factor.
By eqs.(\ref{28}) and (\ref{29}), the thermodynamic bulk pressure
may be written as
\begin{equation}
\kappa\pi = - 3\gamma H^{2} - 2\dot{H}
\ .
\label{30}
\end{equation}
To calculate $\dot{\pi}$ one first has to determine
$\dot{p}$ via
\begin{equation}
\dot{p}  = \frac{\partial p}{\partial n}\dot{n}
+ \frac{\partial p}{\partial T} \dot{T}
\ .  \label{31}
\end{equation}
Using the balance law (\ref{13}) for $n$,  and
the evolution equation (\ref{23}) for
$T$, we find
\begin{equation}
\dot{p}  = - \left(3H - \Gamma\right)
\left(\rho + p\right)c_{s}^{2}
- 3H\pi\left(\frac{\partial_T p}{\partial_T\rho}\right)
\ ,  \label{32}
\end{equation}
with the adiabatic sound velocity $c_{s}$ given by
\begin{equation}
c_{s}^{2} =
\left(\frac{\partial p}{\partial \rho}\right)_{isentropic}
= \frac{n}{(\rho + p)}
\frac{\partial p}{\partial n} + \frac{T}{(\rho + p)}
\frac{\left(\partial_{T} p\right)^{2}}
{\partial_{T} \rho}
 \ .\label{33}
\end{equation}
By differentiating (\ref{29}), we get
\begin{eqnarray}
\dot{\pi} &=& - \frac{2}{3}\frac{\ddot{H}}{H^{2}} \rho
- 2\frac{\dot{H}}{H} \rho
\left(1 +
\frac{\partial_T p}
{\partial_T \rho}\right)
- 3H\left(\rho + p\right)
\frac{\partial_T p}
{\partial_T \rho}\nonumber\\ &&
{}+ \left(3H - \Gamma\right)
\left(\rho + p\right)c_{s}^{2}
\ .  \label{34}
\end{eqnarray}
Using (\ref{30}) and (\ref{34}) in (\ref{27}),
the evolution equation for $H$ becomes
\begin{eqnarray}
\tau\left[\frac{\ddot{H}}{H^{2}}
+ 3\frac{\dot{H}}{H}
\left(1 +
\frac{\partial_T p}
{\partial_T \rho}\right)
+ \frac{9}{2}H \gamma \left(
\frac{\partial_T p}
{\partial_T \rho}
- c_{s}^{2}-c_b^2\right)
\right.&&\nonumber\\
\left.{} + \frac{3}{2}c_{s}^{2}\gamma \Gamma
\right]
+ \frac{\dot{H}}{H^{2}} + \frac{3}{2}\gamma  = 0 \ . &&
\label{35}
\end{eqnarray}

Equation (\ref{35}) is the causal dynamical equation in general
form. We now specify the equations of state implicit in $\rho$
and $p$.
We assume that fluid $1$ is described by the equations of state for 
nonrelativistic matter, i.e.
\begin{equation}
\rho _{1} = n _{1} m + {\textstyle{3\over2}}
n _{1} T \ , \quad
p _{1} = n _{1} T \ , \quad m \gg T \ ,
\label{36}
\end{equation}
while fluid $2$ is relativistic:
\begin{equation}
\rho _{2} = 3 n _{2}T \ ,\quad
p _{2} = n _{2} T  \ .
\label{37}
\end{equation}
Consequently (\ref{35}) reduces to
\begin{eqnarray}
\tau\left[\frac{\ddot{H}}{H^{2}}
+ \left(\frac{5 n _{1} + 8 n _{2}}{n _{1} + 2n _{2}}\right)
\frac{\dot{H}}{H}
+3\left\{
\frac{(n_1 + n_2)\, n_1 m}{(n_1 + 2n_2) \,(n_1 m + 3n_2T)}
- {3\over2}\gamma c_b^2   \right\} H
\right.&&\nonumber\\
\left.{} +
 \frac{(5 n _{1} + 8 n _{2})n_1m}{8(n _{1} + 2n _{2})
(n _{1}m + 3 n _{2} T)} Q \right]
+ \frac{\dot{H}}{H^{2}} +
\frac{3(n _{1}m + 4 n _{2} T)}{2(n _{1}m + 3 n _{2} T)}
 = 0 \ , &&
\label{38}
\end{eqnarray}
where $Q \equiv |\Gamma _{1}| > 0$
was assumed, i.e. the nonrelativistic component decays.
This is the causal evolution equation to be solved.

\section{Causal dynamics of reheating}

The equation (\ref{38}) is a complicated
nonlinear second order equation.
To solve it, one could resort to numerical integration.
However, this is strongly dependent on initial conditions
and on the form of the decay rate $Q$, and it
cannot readily give an idea of the overall dynamical features
implied by the equation. Furthermore, part of the purpose of our
thermodynamical approach is to avoid detailed complexities and to
aim for an overall qualitative understanding
arising from the constraint of causality.

In Section IV.A,
we use eq.(\ref{38}) to estimate the temperature and
decay rate during reheating, neglecting the small perturbation
of the expansion rate due to thermodynamic viscous effects.
(Notice that the gravitational field equations `feel' only the
thermodynamic viscous pressure.)
It turns out that under this condition the reheating dynamics may
be discussed in terms of
the ratio $c _{b}/c _{s}$ of the dissipative to the
adiabatic contributions to the speed of sound.
Then in Section IV.B,
we calculate the perturbation of the expansion rate
in a simple case.

\subsection{Temperature and decay rate}

It is reasonable to assume that the expansion
is approximately governed by
non-viscous effects, and that the latter
can  be treated as a back-reaction.
In this approximation, we neglect the terms multiplying $\tau$
in (\ref{38}), and arrive at the equation governing the expansion
rate:
\begin{equation}
\frac{\dot{H}}{H^2} + \frac{3(n _{1}m + 4 n_{2}T)}
{2(n_{1}m + 3 n_{2}T)} \approx  0
\ .
\label{39}
\end{equation}
We solve this for three different stages of the decay.
Then we use (\ref{38}) to determine the
corresponding decay rate $Q$. The evolution of the temperature
is given by
\begin{equation}
{\dot{T}\over T} = \frac{1}{6(n_1 + 2n_2)T}\left[ -12 H
(p_1 + p_2 + \pi) + n_1 m Q \right]
\label{41}
\end{equation}
as follows from (\ref{24b})  and from
\[
\Gamma = \frac{n_1}{4 n}\frac{m}{T} Q ,
\]
which is a consequence of (\ref{18})
and the equations
of state (\ref{36}) and (\ref{37}).
In the perfect fluid limit $\pi = Q = 0 $ one recovers
$\dot{T}/T = - 2 H$ for $n _{1} \gg n _{2}$, i.e. the
nonrelativistic case, while for radiation ($n _{1} \ll n _{2}$) the  
well--known behaviour
$\dot{T}/T = - H$ is reproduced.
Assuming that $|\pi| \ll p_1 + p_2$, in agreement with the
applicability conditions of the Isreal-Stewart theory,
equation (\ref{41}) may be
written as
\begin{equation}
{\dot{T}\over T} \approx - 2 H\left(
\frac{n _{1} + n _{2}}{n _{1} + 2 n _{2}}\right)
\left[1 - \frac{\Gamma }{3 H}\right]\ .
\label{410}
\end{equation}
It follows that the temperature rises when $\Gamma>3H$. By (\ref{13}),
this is equivalent to $\dot{n}>0$, i.e. to a growth in the total
particle number density (cf. \cite{ZPMN}).
As discussed in Section I, such abundant
net particle creation is expected to occur in the initial
nonrelativistic stage of reheating.
\[ \]
\noindent (i) \underline{Nonrelativistic regime}\\

In this regime the fluid is dominated by massive particles, so that
$n_1 \gg n_2$, $\gamma \approx  1$. Then
(\ref{39}) reduces to
\begin{equation}
\dot{H} + {\textstyle{3\over2}}H^2
\approx  0 \quad\Rightarrow\quad H \approx  \frac{2}{3t}
\ .
\label{40}
\end{equation}
Inserting this solution back into (\ref{38}), we find that the decay 
rate at the beginning of reheating is
\begin{equation}
Q \approx  {\textstyle{36\over5}} c_b^2  H
\quad\Rightarrow\quad \Gamma \approx 3 \left(
\frac{c _{b}}{c _{s}}\right)^2H \quad\mbox{where}\quad
c _{s}^{2} \approx \frac{5}{3}\frac{T}{m}
\ .
\label{40a}\end{equation}
It is a specific feature of the present first-order approximation
that it fixes the rates $Q$ and $\Gamma $ which, in our general
setting, are input parameters. $Q$ and $\Gamma $ are positive
and proportional to $H$.
Moreover, our approximation relates $c _{b}$, the bulk viscous
contribution to the sound speed, to $Q$ and $\Gamma $.
The exact value of $\Gamma $ is fixed by the ratio
$c _{b}^{2}/c _{s}^{2}$. Since, according to (\ref{40a}),
$c _{s}^{2} \ll 1$,
a value $c _{b}^{2} \approx 1$ is admitted by (\ref{26}).
This amounts to $\Gamma \gg H$, i.e., violent particle
production as required for `preheating'.
A sufficiently high value of $\Gamma $ also determines the entropy  
production.
Neglecting second-order terms in $\pi $, the entropy flow vector is  
given
by $S ^{a} \approx n s u ^{a}$, where $s$ is the entropy per particle. 
For large $\Gamma $, neglecting the change in the entropy per
particle, the entropy production density is approximately
\begin{equation}
S ^{a}_{;a} \approx   3\left(
\frac{ c _{b}}{c _{s}}\right)^2 s H
\ .
\label{40b}
\end{equation}
Equation (\ref{41}) implies that at the beginning of the
reheating, the rate of temperature change is given by
\begin{equation}
{\dot{T}\over T} \approx - 2 \left[1 -\left(
\frac{c _{b}}{c _{s}}\right)^2 \right] H
\,.
\label{41a}\end{equation}
In the initial stage, a very large rate of creation
of particles can lead
to a growth in the net number density and thus in the temperature.
By (\ref{41a}), this is again equivalent to the effective dissipative
contribution $c_b$ to the  sound speed exceeding the adiabatic
contribution $c_s$:
\begin{equation}
\dot{n}>0\quad\Leftrightarrow\quad\dot{T}>0\quad
\Leftrightarrow\quad c_b>c_s
\label{41c}\end{equation}
Equation (\ref{41a}) then implies that
initially the temperature rises
extremely rapidly (see also \cite{ZPMN}). It
reaches a maximum, which
we could call the reheating temperature $T_{reh}$, following standard
terminology, and then decreases as $c_b$ falls below $c_s$.
By (\ref{40a}), the reheating temperature is given by
\begin{equation}
T_{reh}\approx {\textstyle{3\over5}}c_b^2 m
\ .
\label{41b}\end{equation}
It follows from (\ref{40b}) that there is high entropy production
in this regime.
Alternatively, this regime may be characterized in terms of the
effective bulk pressure (\ref{24a}).
For $\Gamma \geq 3 H $ the condition
$|\pi| \ll p _{1} + p _{2} = p$ leads generally to
\begin{equation}
\pi _{_{eff}} \approx - \frac{\Gamma }{3 H} p\ .
\label{41d}
\end{equation}
Using the relations (\ref{40a}) one gets
\begin{equation}
\pi _{_{eff}} \approx - \left(\frac{c _{b}}{c _{s}}\right)^{2} p
\approx - \frac{3}{5}n  m c _{b}^{2}
 \approx - n T _{reh}\ ,
\label{41e}
\end{equation}
equivalent to
\begin{equation}
\frac{|\pi _{_{eff}}|}{p} \approx \frac{T _{reh}}{T}\ .
\label{41f}
\end{equation}
For large particle production the condition $|\pi _{_{eff}}| \gg p$  
is fulfilled.
The reheating temperature (\ref{41b}) may be understood as
the temperature for which
$|\pi _{_{eff}}| \approx p \approx p _{1}$.

\[ \]
\noindent (ii) \underline{Intermediate regime}\\

Here the energy densities of both components are comparable,
i.e. $ n_1 m \approx 3 n_2 T$ so that $\gamma\approx{7\over6}$ and  
consequently equation (\ref{39})
simplifies to
\begin{equation}
\dot{H} + {\textstyle{7\over4}}H^2 \approx 0  \quad \Rightarrow
\quad H \approx  \frac{4}{7t} \ . \label{42}
\end{equation}
Then (\ref{38}) implies that the decay rate is
\begin{equation}
Q \approx  \frac{1}{2}\left(1 + 42 c _{b}^{2} \right) H
\quad \Rightarrow
\quad \Gamma \approx  \frac{3}{8}
\left[1 + 8\left( \frac{c _{b}}{c _{s}}\right)^2 \right] H \quad
\mbox{where}\quad
c _{s}^{2}  \approx \frac{4}{21}
\ .
\label{42a}\end{equation}
As can be seen from (\ref{41}), the temperature
behaviour is
\begin{equation}
\frac{\dot{T}}{T} \approx  -{7\over8}
\left[1 - \frac{8}{7}\left(\frac{c _{b}}{c _{s}}\right)^2\right]H
\ .
\label{42b}\end{equation}
After the initial stage in the nonrelativistic regime,
the creation of ultrarelativistic particles slows down, while
the nonrelativistic particles continue to decay,  and we no longer
expect that $\dot{n}$ is positive.
Thus $\dot{T}<0$, although by (\ref{42b}) the cooling rate
is less than the non--dissipative case ($c_b=0$).
We expect that the temperature should decrease monotonically after  
reaching its maximum $T _{reh}$.
By (\ref{42b}) this is the case provided that
\begin{equation}
c _{b}^{2} < {\textstyle{7\over8}}c_s^2\quad\Rightarrow\quad
c_b^2<{\textstyle{1\over6}}\,.
\label{42c}\end{equation}
We expect that (\ref{42c}) is easily satisfied in the intermediate
regime, after the initial violent rate of
creation of particles
has passed, and the two--fluid mixture evolves increasingly towards
`normal' two--fluid behaviour, for which $c_b\ll 1$ (see equation (53)
of \cite{ZMN}).
\[ \]
\noindent (iii) \underline{Ultra-relativistic regime}\\

In this last stage of reheating, the energy density
becomes dominated by
the radiation fluid, i.e. $n_2T \gg n_1m$ and
$\gamma \approx {4\over3}$. Therefore from (\ref{39}) we have
\begin{equation}
\dot{H} + 2H^2 \approx  0  \quad \Rightarrow \quad
H \approx  \frac{1}{2t} \ .
\label{43}
\end{equation}
Then (\ref{38}) implies the decay rate:
\begin{equation}
Q \approx  36 c _{b}^{2}\left( \frac{n_2 T}{n_1 m}\right)H
\quad \Rightarrow
\quad \Gamma \approx 3\left( \frac{c _{b}}{c _{s}}\right)^2 H\quad
\mbox{where}\quad
 c _{s}^{2} \approx \frac{1}{3}\ .
\label{43a}\end{equation}
The temperature change
follows from (\ref{41}) as
\begin{equation}
\frac{\dot{T}}{T} \approx -
\left[1 -  \left(\frac{c _{b}}{c _{s}}\right)^2 \right] H
\ .
\label{43b}\end{equation}
As argued above, we expect that $c_b<c_s$ is easily satisfied, so
that the temperature continues to fall, although still at a
reduced rate relative to the non--dissipative case.
Towards the end of reheating the cosmic medium  approaches a
perfect, relativistic fluid with vanishing $\zeta $, so that $c
_{b}^{2} $ must tend to zero.
For any nonzero $c _{b}^{2}$ the decay rate $Q$ diverges in
the limit
$n_1 m \ll n_2 T$.

\subsection{Perturbations of the expansion rate}

It is possible, given {\it a priori} the forms
of the relaxation time $\tau$ and the decay rate $Q$, to calculate
the perturbations of the expansion rate $H$ due to causal
viscous and reaction effects, via (\ref{38}). We illustrate this
with a simple model, which in particular can accomodate oscillations
in $H$ during the initial stage of
reheating (compare \cite{mfb}, p241).

The simple model is based on the ansatz that $\tau^{-1}$ and $Q$ are
proportional to the expansion rate in the initial stage of reheating, 
i.e.
\begin{equation}
\tau^{-1}=\nu H \ , \quad Q=\beta H \ ,
\label{50}\end{equation}
where $\nu$ and $\beta$ are positive constants (with $\nu>1$ for
a consistent hydrodynamic description). The ansatz for $Q$ is
consistent with (\ref{40a}) if $c_b$ is constant.
Using (\ref{50}) in (\ref{38}) in
the nonrelativistic regime, we find that the evolution
of the expansion rate is governed by
\begin{equation}
\ddot{H}+(5+\nu)H\dot{H}+\left[3+{\textstyle{3\over2}}\nu+
{\textstyle{5\over8}}\beta-{\textstyle{9\over2}}c_b^2\right]H^3
\approx0 \ .
\label{51}\end{equation}
Now we know that
\begin{equation}
H={2\over3t}+h\quad\mbox{where}\quad |h|\ll {2\over3t} \ .
\label{52}\end{equation}
Substituting (\ref{52}) into (\ref{51}) and linearising,
we find that the perturbation $h$ is governed by
\begin{eqnarray*}
\ddot{h}+\left[{\textstyle{2\over3}}(5+\nu)\right]t^{-1}\dot{h}
+\left[{\textstyle{1\over6}}\left(4+8\nu+5\beta-36c_b^2\right)\right]
t^{-2}h && \\
{}\approx\left[{\textstyle{1\over27}}\left(36c_b^2-5\beta\right)\right]
t^{-3}\ . &&
\end{eqnarray*}
This is the equation of a forced damped oscillator, as is readily seen
after the change of variable to $s=\ln t$:
\begin{eqnarray}
h''+\left[{\textstyle{1\over3}}(7+2\nu)\right]h'
+\left[{\textstyle{1\over6}}\left(4+8\nu+5\beta-36c_b^2\right)\right]
h && \nonumber\\
{}\approx\left[{\textstyle{1\over27}}
\left(36c_b^2-5\beta\right)\right]e^{-s} \ . &&
\label{54}\end{eqnarray}

The simplest case is (\ref{40a}), i.e.,
$5 \beta = 36 c _{b}^{2}$, for which (\ref{54})
leads to the overdamped perturbation
\begin{equation}
h(t)\approx\epsilon_1t^{-2}+\epsilon_2t^{-(1+2\nu)/3}\ .
\label{55}\end{equation}
Damped oscillations of $h$ about the zero--order solution $2/(3t)$
occur when
\begin{equation}
30\beta > 216 c_b^2+(2\nu-5)^2 \ ,
\label{56}\end{equation}
which shows that $c_b$ is a damping factor, while the decay
coefficient $\beta$ contributes to oscillation. A further constraint
on the thermodynamic parameters arises from the requirement that the
particular integral of (\ref{54}) must be small compared to
$2/(3t)$:
\begin{equation}
|B|\equiv
\left|{36c_b^2-5\beta \over 5\beta-36c_b^2+4\nu-4}\right|\ll 1 \ .
\label{57}\end{equation}
If (\ref{56}) and (\ref{57}) are satisfied, then the damped
oscillatory perturbation is given by
\begin{equation}
h(t)\approx{\textstyle{2\over9}}Bt^{-1}+t^{-(7+2\nu)/6}
\left[\epsilon_1\cos(\omega\ln t)+\epsilon_2\sin(\omega\ln t)\right]\ ,
\label{58}\end{equation}
where
$$
\omega={\textstyle{1\over6}}\left[30\beta-216c_b^2-(2\nu-5)^2
\right]^{1/2}
$$
is the frequency of oscillation. A simple choice that satisfies
(\ref{56}) and (\ref{57}) is
$$
\nu={\textstyle{5\over2}}\ ,\quad\beta={\textstyle{2\over165}}+
{\textstyle{36\over5}}c_b^2 \ ,
$$
in which case
$$
H\approx{2\over3t}-{1\over450t}+{1\over t^2}\epsilon\exp\left(i
{\sqrt{11}\over33}\ln t\right)
\ .
$$

Finally, we note that the perturbations of the Hubble rate induce
perturbations of the temperature via the temperature evolution
equation (\ref{41}). Using (\ref{50}) and $n_1\gg n_2$,
$p_1+p_2+\pi\approx n_1 T$, we find
$$
\dot{T}+2HT\approx {\textstyle{1\over6}}\beta mH \ .
$$
Denoting by $\bar{T}$ the dominant zero--order part of the
temperature, which is determined by $\bar{H}\equiv 2/(3t)$,
and by $\delta T$ the perturbation induced by $\delta H\equiv h$,
this equation leads to
\begin{equation}
(\delta T)^{\displaystyle{\cdot}}+2\bar{H}\,\delta T\approx
\left({\textstyle{1\over6}}\beta m-2\bar{T}\right)h
\ .
\label{59}\end{equation}
Equation (\ref{59}) determines the temperature perturbation
explicitly in terms of the Hubble rate perturbation. Clearly, if
$h$ has an oscillatory component, then so will $\delta T$.

\section{Concluding remarks}

A qualitative analysis based on a thermodynamic two--fluid
model subject to causality has produced overall features which are
consistent with expectations. The decay rate of nonrelativistic
particles implied by causality
is positive throughout the reheating process (thus providing
a consistency check on our approach), and proportional to
the expansion. The overall behaviour of the temperature and decay
rate are essentially determined by
$c_b$, the dissipative contribution to the sound speed. Provided that
$c_b$ exceeds the adiabatic contribution $c_s$ in the beginning of
reheating, equivalently, provided that the total number density grows
by virtue of super--abundant particle creation,
the temperature rises very
rapidly at the start of reheating. It quickly reaches a maximum
value (\ref{41b}), whereafter it
falls with expansion, at a reduced rate (determined by $c_b$) relative
to the non--dissipative case.
Large amounts of entropy are generated in the early stage, as shown by 
(\ref{40b}).
A simplified model (\ref{50})
allows us to calculate explicitly the decaying perturbations of the
Hubble rate, including a case of oscillatory perturbations (\ref{58})
around matter--dominated behavior.

We have thus shown
the capacity of a phenomenological model based on
causal relativistic
thermodynamics to predict the expected
basic features of reheating with economy and simplicity, and
independent of detailed knowledge of the interaction.
This should
be counted as another success of the Israel--Stewart theory (as
adapted to deal with interacting fluids), and another argument
in favour of the theory with its causality and stability properties.

\acknowledgments

This work has been partially supported by the Deutsche
Forschungsgemeinschaft, the Spanish Ministry of Education
(grant PB94-0718), NATO (grant CRG 940598) and grants from
Portsmouth and Natal Universities.
WZ thanks the Grup de F\'{\i}sica Estad\'{\i}stica,
Universitat Aut\`{o}noma de Barcelona for
warm hospitality.  DP is grateful to the
Department of Theoretical Physics of
the University of Konstanz,
where part of this work was done.
RM thanks the Department of Physics at Universitat Aut\`{o}noma
de Barcelona and the Department of Mathematics at the University
of Natal, for warm hospitality.
WZ and DP would like to thank Esteban Calzetta for discussions.

\end{document}